\documentclass[10pt,conference]{IEEEtran}
\usepackage[utf8]{inputenc}
\usepackage[T1]{fontenc}
\usepackage{microtype}
\usepackage{flushend}

\usepackage{booktabs}
\usepackage{listings}

\usepackage{fancybox}
\usepackage{adjustbox}
\usepackage{subfigure}

\usepackage{xcolor}
\usepackage{cite}
\usepackage{listings}
\usepackage{hyperref}
\usepackage{amsmath,amssymb,amsfonts}

\newcommand{\code}[1]{\sloppy{\ttfamily #1}}

\newenvironment{resultbox}
{	\noindent
	\begin{center}
		\begin{Sbox}
			\begin{minipage}{.9\linewidth}
				\smallskip\indent
			}
			{                               %\smallskip
			\end{minipage}
		\end{Sbox}\fbox{\TheSbox}
	\end{center}
}

\definecolor{pblue}{rgb}{0.13,0.13,1}
\definecolor{pgreen}{rgb}{0,0.5,0}
\definecolor{pred}{rgb}{0.9,0,0}
\definecolor{pgrey}{rgb}{0.46,0.45,0.48}
\definecolor{bg}{rgb}{0.95,0.95,0.95}
\newtheorem{definition}{Definition}

\newcommand{\topWarnings}{20}
\newcommand{\totalWarnings}{280,651}

\newcommand{\expBS}{BS}
\newcommand{\expANNBS}{B\textsubscript{ANN}S}
\newcommand{\expSS}{SS}
\newcommand{\expBB}{BB}

\begin{document}

\title{Neural Bug Finding:\\ A Study of Opportunities and Challenges}

\author{
\IEEEauthorblockN{Andrew Habib}
\IEEEauthorblockA{
	\textit{Department of Computer Science}\\
	\textit{TU Darmstadt}\\
	Germany\\
	andrew.a.habib@gmail.com
}
\and
\IEEEauthorblockN{Michael Pradel}
\IEEEauthorblockA{
	\textit{Department of Computer Science}\\
	\textit{TU Darmstadt}\\
	Germany\\
	michael@binaervarianz.de
}
}

\maketitle

\begin{abstract}

Static analysis is one of the most widely adopted techniques to find software bugs before code is put in production. 
Designing and implementing effective and efficient static analyses is difficult and requires high expertise, which results in only a few experts able to write such analyses.
This paper explores the opportunities and challenges of an alternative way of creating static bug detectors: neural bug finding.
The basic idea is to formulate bug detection as a classification problem, and to address this problem with neural networks trained on examples of buggy and non-buggy code.
We systematically study the effectiveness of this approach based on code examples labeled by a state-of-the-art, static bug detector.
Our results show that neural bug finding is surprisingly effective for some bug patterns, sometimes reaching a precision and recall of over 80\%, but also that it struggles to understand some program properties obvious to a traditional analysis.
A qualitative analysis of the results provides insights into why neural bug finders sometimes work and sometimes do not work.
We also identify pitfalls in selecting the code examples used to train and validate neural bug finders, and propose an algorithm for selecting effective training data.
\end{abstract}

\section{Introduction}
\label{sec:intro}

%Software bugs remain one of the key challenges for developers, because they cause immense costs, may have disastrous consequences, and degrade user experience.
A popular way of finding software bugs early during the development process is static analysis tools that search a code base for instances of common bug patterns.
These tools, which we here call \emph{bug detectors}, often consist of a scalable static analysis framework and an extensible set of checkers that each search for instances of a specific bug pattern.
Examples of bug detectors include the pioneering FindBugs tool~\cite{Hovemeyer2004}, its successor SpotBugs\footnote{\url{https://spotbugs.github.io/}}, Google's Error Prone tool~\cite{Aftandilian2012}, and the Infer tool by Facebook~\cite{Calcagno2015}.
%Many organizations deploy such tools on a regular basis, e.g., as part of their process for committing code changes to the main code base.

Despite the overall success of static bug detection tools, there still remains a lot of potential for improvement.
A recent study that applied state-of-the-art bug detectors to a set of almost 600 real-world bugs shows that over 95\% of the bugs are currently missed~\cite{ase2018-study}.
The main reason is that various different bug patterns exist, each of which needs a different bug detector.
These bug detectors must be manually created, typically by program analysis experts, and they require significant fine-tuning to find actual bugs without overwhelming developers with spurious warnings.
Bug detectors often require hundreds of lines of code each, even for bug patterns that seem trivial to find at first sight and when being built on top of a static analysis framework.

This paper studies a novel way of creating bug detectors: \emph{neural bug finding}.
Motivated by the huge success of neural networks for various software engineering tasks~\cite{Allamanis2018}, we ask a simple question:
Can we automatically learn bug detectors from data, instead of implementing program analyses manually?
Giving a positive answer to this question has the potential of complementing existing bug detectors with additional checkers that address previously ignored bug patterns.
Moreover, it may enable non-experts in program analysis, e.g., ordinary software developers, to contribute to the creation of bug detectors.

Given the importance of bug detection and the power of neural networks, the intersection of these two areas so far has received surprisingly little attention.
Existing work focuses on learning-based defect prediction~\cite{Wang2016a}, which ranks entire files by their probability to contain any kind of bug, whereas we here aim at pinpointing code that suffers from a specific kind of bug.
Other work addresses the problems of predicting code changes~\cite{Yin2018}, predicting identifier names~\cite{Raychev2015,Hellendoorn2018,Context2Name}, and predicting how to complete partial code~\cite{Bruch2009,Raychev2014,Liu2016}, which are complementary to detecting bugs.
The perhaps closest existing work is DeepBugs~\cite{oopsla2018-DeepBugs}, which trains a neural network to find name-related bugs, and learning-based techniques for identifying security vulnerabilities~\cite{Choi2017,Li2018a,Harer2018}.
While these approaches show that neural bug finding is possible for a specific class of bugs, we here study the potential of neural bug finding in much more detail and for a broader range of code issues.

Automatically learning bug detectors requires addressing two problems:
(1) Obtaining sufficient training data, e.g., consisting of buggy and non-buggy code examples.
(2) Training a model that identifies bugs, e.g., by distinguishing buggy code from non-buggy code.
The first problem could be addressed by automatically seeding bugs into code, by extracting buggy code examples from version histories, or by manually labeling code examples as instances of specific bug patterns.
In this work, we sidestep the first problem and study whether given sufficient training data, the second problem is tractable.
Our work therefore does not yield a ready-to-deploy bug detection tool, but rather novel insights into what kinds of bugs neural bug finding can and cannot find.
We believe that thoroughly studying this question in isolation is an important step forward toward the ultimate goal of neural bug finding.

To study the potential of learned bug detectors while sidestepping the problem of obtaining labeled training data, we use an existing, traditionally developed bug detector as a generator of training data.
To this end, we run the existing bug detector on a corpus of code to obtain warnings about specific kinds of bugs.
Using these warnings and their absence as a ground truth, we then train a neural model to distinguish code with a particular kind of warning from code without such a warning.
For example, we train a model that predicts whether a piece of code uses reference equality instead of value equality for comparing objects in Java.
This setup allows us to assess to what extent neural bug finding can imitate existing bug detectors.

One drawback of using an existing bug detector as the data generator is that some warnings may be spurious and that some bugs may be missed.
To mitigate this problem, we focus on bugs flagged by bug detectors that are enabled in production in a major company and that empirically show false positive rates below 10\%~\cite{sadowski2015tricorder}.
Another drawback is that the learned bug detectors are unlikely to outperform the static analyzers they learn from.
However, the purpose of this work is to study whether training a model for neural bug finding is feasible, whereas we leave the problem of obtaining training data beyond existing static analyzers as future work.

The main findings of our study include the following:
\begin{itemize}
	\item Learned bug detectors identify instances of a surprisingly large number of bug patterns with precision and recall over 80\%.
	At the same time, the learned models sometimes fail to understand program properties that a traditional analysis easily finds.
	
	\item Neural bug finding works because the models learn to identify common syntactic patterns correlated with bugs, particular API misuses, or common instances of a more general bug pattern.
	
	\item The composition of the training data a neural bug finding model is learned from has a huge impact on the model's effectiveness.
	We study several strategies for composing training data and present a novel algorithm for selecting effective training examples.
	
	\item More training data yields more effective models, but surprisingly small data sets, e.g., of only 77 examples, can still yield effective neural bug detectors.
	
	\item Following a naive approach for validating a learned bug detector may lead to very misleading results.
	
\end{itemize}

In summary, this paper contributes the first comprehensive study of neural bug finding.
The study reveals novel insights into the opportunities and challenges associated with this novel way of creating bug detectors.
We believe that our work is a step forward toward complementing traditional ways of creating bug detectors.
In particular, the study provides a basis for future work on generating training data for neural bug finding, for developing machine learning models that reason about rich representations of code, and for building neural bug finding tools.
To fuel these and other lines of future work, we make our tool and data set publicly available.\footnote{URL inserted into final version.}

\section{Methodology}
\label{sec:method}

%% MP: omitted, as it repeats the intro
%Ideally, instead of manually writing static analyses to find bugs, we would like to automatically learn bug detectors from examples of buggy and non-buggy code.
%The overall goal of this work is to find out whether deep learning is suitable for and capable of general purpose static bug detection.

\begin{figure}
	\centering
	\includegraphics[width=\linewidth]{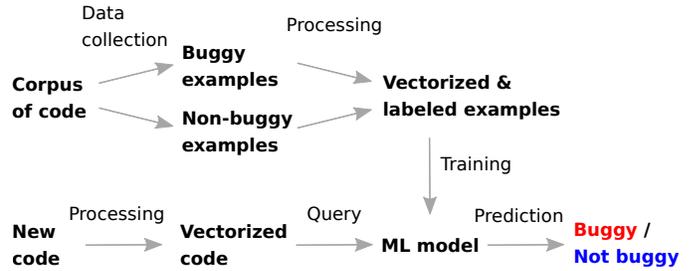}
	\caption{Overview of neural bug finding.}
	\label{fig:approach}
\end{figure}

Our approach applies machine learning (ML), specifically deep learning, to source code and learns a model that predicts whether a given piece of code suffers from a specific bug or not.
\autoref{fig:approach} gives an overview of the neural bug finding approach.
As training and validation data, we gather hundreds of thousands of code examples, some of which are known to contain specific kinds of bugs, from a corpus of real-world software projects (Section~\ref{subsec:collection}).
To feed these code examples into a neural model, we abstract and vectorize the source code of individual methods (Section~\ref{subsec:extraction}).
A particularly interesting challenge is how to select examples of buggy and non-buggy code for training the bug detection model, which we address in Section~\ref{subsec:nonbuggy}.
Finally, Section~\ref{subsec:rnn} describes how we train recurrent neural network (RNN) models that predict different kinds of bugs.

\subsection{Gathering Data}
\label{subsec:collection}

To study the capability of neural bug finding, we need some kind of \emph{oracle} that provides examples of buggy and non-buggy code to train ML models.
One could potentially collect such data from existing bug benchmarks~\cite{Just2014,cifuentes2009begbunch,lu2005bugbench}.
Unfortunately, such bug benchmarks provide at most a few hundreds of buggy code examples, which is a relatively small number for training neural networks.
Other directions include mining existing code repositories for pull requests and commits that fix bugs or generating training data by injecting bugs, e.g., via mutations.

In this work, we obtain examples of buggy and non-buggy code by running a state-of-the-art static analyzer as an oracle on a large corpus of code, and by collecting warnings produced by the static analyzer.
We use Error Prone~\cite{Aftandilian2012} as the oracle, a state-of-the-art static bug finding tool for Java, which is developed and used by Google, and made available as open-source.
We run Error Prone on the Qualitas Corpus~\cite{Tempero2010}, a curated set of 112 open-source Java projects and collect all warnings and errors reported by Error Prone along with their corresponding kinds and code locations.
To simplify terminology, we call all problems reported by Error Prone a ``bug'', irrespective of whether a problem affects the correctness, maintainability, performance, etc.\ of code.

\autoref{table:top-warnings} shows the bug kinds we consider in this work.
Error Prone warnings flag class-level problems, e.g., mutable \code{enum}s; method-level problems, e.g., missing annotations, such as the \code{@Override} annotation (Id~1 in~\autoref{table:top-warnings}); and statement-level and expression-level issues, such as expressions with confusing operator precedence (Id~9 in~\autoref{table:top-warnings}).
Since most of the warnings are at the method level or at the expression level, our study focuses on learning to predict those bugs, ignoring class-level bugs.
After removing class-level bugs, \autoref{table:top-warnings} includes the 20 most common kinds of bugs reported by Error Prone on the Qualitas corpus.

To illustrate that finding these bugs with traditional means is non-trivial, the last column of \autoref{table:top-warnings} shows how many non-comment, non-empty lines of Java code each bug detector has.
On average, each bug detector has 170 lines of code, in addition to th  156k lines of general infrastructure and test code in the Error Prone project.
These numbers show that manually creating bug detectors is a non-trivial effort that would be worthwhile to complement with learned bug detectors.

\begin{table*}
	\centering
	\caption{Top \topWarnings{} warnings reported by Error Prone on the Qualitas Corpus.}
	\label{table:top-warnings}
	\begin{adjustbox}{center,width=0.9\textwidth}
	\begin{tabular}{rlrlr}
		\toprule
		 \textbf{Id} & \textbf{Warning} & \textbf{Count} & \textbf{Description} & \textbf{LoC} \\
		 \midrule
		1 & MissingOverride & 268,304 & Expected \code{@Override} because method overrides method in supertype; including interfaces & 111 \\
		2 & BoxedPrimitiveConstructor & 3,769 & \code{valueOf} or autoboxing provides better time and space performance & 268 \\
		3 & SynchronizeOnNonFinalField & 2,282 & Synchronizing on non-final fields is not safe if the field is updated & 66 \\
		4 & ReferenceEquality & 1,680 & Compare reference types using reference equality instead of value equality & 282 \\
		5 & DefaultCharset & 1,550 & Implicit use of the platform default charset, can result in unexpected behaviour & 515 \\
		6 & EqualsHashCode & 590 & Classes that override equals should also override hashCode & 106 \\
		7 & UnsynchronizedOverridesSynchronized & 517 & Thread-safe methods should not be overridden by methods that are not thread-safe & 125 \\
		8 & ClassNewInstance & 486 & \code{Class.newInstance()} bypasses exception checking & 254 \\
		9 & OperatorPrecedence & 362 & Ambiguous expressions due to unclear precedence & 118 \\
		10 & DoubleCheckedLocking & 204 & Double-checked locking on non-volatile fields is unsafe & 305 \\
		11 & NonOverridingEquals & 165 & A method that looks like \code{Object.equals} but does not actually override it & 179\\
		12 & NarrowingCompoundAssignment & 158 & Compound assignments like \code{x += y} may hide dangerous casts & 167\\
		13 & ShortCircuitBoolean & 116 & Prefer the short-circuiting boolean operators \code{\&\&} and \code{||} to \code{\&} and \code{|} & 88 \\
		14 & IntLongMath & 111 & Expression of type int may overflow before being assigned to a long & 127 \\
		15 & NonAtomicVolatileUpdate & 80 & Update of a volatile variable is non-atomic & 142 \\
		16 & WaitNotInLoop & 77 & \code{Object.wait()} and \code{Condition.await()} must be called in a loop to avoid spurious wakeups & 76\\
		17 & ArrayToString & 56 & Calling \code{toString} on an array does not provide useful information (prints its identity) & 256 \\
		18 & MissingCasesInEnumSwitch & 53 & Switches on \code{enum} types should either handle all values, or have a default case & 86 \\
		19 & TypeParameterUnusedInFormals & 46 & A method’s type parameter is not referenced in the declaration of any of the formal parameters & 135 \\
		20 & FallThrough & 45 & \code{switch} case may fall through & 96 \\
		\midrule
		& \textbf{Total} & \textbf{\totalWarnings{}} && \textbf{3,402} \\
		\bottomrule
	\end{tabular}
	\end{adjustbox}
\end{table*}

\subsection{Representing Methods as Vectors}
\label{subsec:extraction}

\subsubsection{Code as Token Sequences}

The next step is modeling source code in a manner that allows us to apply machine learning to it to learn patterns of buggy and non-buggy code.
Among the different approaches, we here choose to represent code as a sequence of tokens.
This representation is similar to natural languages~\cite{Hindle2012,collobert2011natural}
and has seen various applications in programming and software engineering tasks, such as bug detection~\cite{Wang2016}, program repair~\cite{Bhatia2016,Gupta2017}, and code completion~\cite{Raychev2014}.

Let $M$ be the set of all non-abstract Java methods in our corpus of code.
For each method $m \in M$, we extract the sequence of tokens $s_m$ from the method body, starting at the method definition and up to length $n$.
Let $S$ be the set of all sequences extracted from all methods $M$.
Extracted tokens include keywords such as \code{for}, \code{if}, and \code{void}; separators such as \code{;}, \code{()}, and \code{,}; identifiers such as variable, method, and class names; and finally, literals such as \code{5} and \code{"abc"}.
Each token $t_{i} = (lex, t, l)$,  where $1 \leq i \leq n$, is a tuple of the lexeme itself, its type $t$, and the line number $l$ at which $t$ occurs in the source file.
We ignore comments.
As a default, we choose a sequence length of $n = 50$ in our experiments.

As an alternative to a token sequence-based code representation, we could model code, e.g., as abstract syntax trees (ASTs), control-flow graphs (CFGs), or program-dependence graphs (PDGs).
Recent work has started to explore the potential of graph-based code representations~\cite{DeFreez2018,Allamanis2017b,Brockschmidt2018,Alon2018,Alon2018a}.
We here deliberately focus on a simpler, sequence-based code representation, so that our study provides a lower bound on the potential effectiveness of neural bug finding, leaving the use of richer code representations as future work.

\subsubsection{Representing Tokens}

To enable the ML model to learn and generalize patterns from source code, we abstract the extracted token sequences in such a way that discovered patterns are reusable across different pieces of code.
One challenge is that source code has a huge vocabulary due to identifiers and literals chosen by developers~\cite{Babii2019}.
To mitigate this problem, we extract a vocabulary $V$ consisting of the most frequent keywords, separators, identifiers, and literals from all code in our corpus.
In addition to the tokens in the corpus, we include two special tokens: \code{UNK}, to represent any occurrence of a token beyond the most frequent tokens, and \code{PAD} to pad short sequences.
%total number of words in vocab: 952,265
%Total tokens occuerences: 57,217,223
%Total tokens covered by top 998 vocab: 46,790,448
% % of covered tokens: 0.82
In our experiments, we set $|V| = 1000$ which covers $82\%$ of all keywords, separators, identifiers, and literals in our corpus.

%Since ML models are probabilistic models which operate on real-valued vectors, we need to represent source code as vectors of real numbers.
%To achieve this, we have to learn real-valued vector representations, called \emph{embeddings}, of individual tokens in the source code.
%
%\begin{definition}[Token Embeddings]
%	Token embedding is a function $E: t \rightarrow \mathbb{R}^e$ which maps a token $t$ to a real-valued vector of dimension $e$.
%\end{definition}
%
%Embeddings could be learned in different ways.
%One approach to learning embeddings is to use pre-trained Word2Vec models from natural language data sets~\cite{Mikolov2013a}.
%Another approach is to pre-train the embeddings in an unsupervised manner, for instance, using the surrounding context of the extracted tokens~\cite{oopsla2018-DeepBugs}.
%
%In this work, we learn the embeddings while training the neural model.
%We first use one-hot vector representation of each word in the vocabulary set $V$ such that each token vector is all zeros and only one specific entry is set to represent a specific token.
%Then, using an embedding layer of size $e$, our neural bug finding model learns the tokens embeddings while training the neural network to distinguish buggy and non-buggy code examples.
%We set $e=50$ in our experiments.

We convert the sequences of tokens of a given code example to a real-valued vector by representing each token $t$ through its one-hot encoding.
The one-hot encoding function $H(t)$ returns a vector of length $|V|$, where all elements are zero except one that represents the specific token $t$.
To allow the learned models to generalize across similar tokens, we furthermore learn an embedding function $E$ that maps $H(t)$ to $\mathbb{R}^e$, where $e$ is the embedding size.
Based on these two functions, we represent a sequence of tokens $s \in S$ through a real-valued vector $v_s$ as follows:

\begin{definition}[Source Code Vector]
	
	For a sequence of tokens $s \in S$ of length $n$, where $s = t_1, t_2, \dots, t_n$ is extracted from a source code method $m \in M$,
	the vector representation of $s$ is $v_s = [E(H(t_1)), E(H(t_2)), \dots, E(H(t_n))]$.
	\label{def:vec}
\end{definition}

%\todo{Are the token embeddings learned jointly with the models, or pre-trained?}

\subsection{Buggy and Non-Buggy Examples}
\label{subsec:nonbuggy}

The training and validation data consists of two kinds of code examples: buggy and non-buggy examples.
We focus on methods as code examples, i.e., our neural bug detectors predict whether a method contains a particular kind of bug.
Let $K$ be the set of all bug kinds that the oracle can detect and $W$ be the set of all warnings reported by it on the Qualitas corpus.
Each warning $w \in W$ is represented as $w = (k, l, m)$ where $k \in K$ is the bug kind flagged at line number $l$ in method $m$.
%Let $M$ be the set of all non-abstract Java methods in our corpus of code.
For each kind of bug $k \in K$, we consider two subsets of $M$:
\begin{itemize}
	\item The set $M_{k_{\mathrm{bug}}}$ of methods flagged by the oracle to suffer from bug kind $k$.
	\item The set $M_{k_{\mathrm{nBug}}}$ of methods for which the oracle does not report any bug of kind $k$.
\end{itemize}

Based on these two sets, we select a subset of the methods as examples to train and validate our models, as described in the following.
After selecting the methods, we produce two sets of sequences, $S_{k_{\mathrm{bug}}}$ and $S_{k_{\mathrm{nBug}}}$, as described in Section~\ref{subsec:extraction}.

\subsubsection{Selecting Non-Buggy Examples}
\label{subsubsec:select_nonbuggy}
One strategy for selecting non-buggy examples is to randomly sample from all methods that are not flagged as buggy for a bug of kind $k$.
However, we found this naive approach to bias the learned model towards the presence or absence of specific tokens related to $k$, 
%instead of learning the underlying bug pattern.
but not necessarily sufficient to precisely detect $k$.
For example, when training a model to predict a problem with binary expressions (Id~9 in~\autoref{table:top-warnings}), using the naive approach to select non-buggy examples would result in a model that learns to distinguish source code sequences that contain binary expressions from sequences that do not.
In other words, it would simply flag any binary expression as potentially buggy.

To address this problem, we selectively pick non-buggy examples that are similar to the buggy examples, but that do not suffer from the same programming error $k$.
For example, if a warning kind $k$ flags binary expressions, we would like $S_{k_{\mathrm{nBug}}}$ to be mostly composed of sequences that include binary expressions but that do not suffer from $k$.
To select such similar examples in an automated manner, we perform two steps.
First, we convert each sequence into a more compact vector that summarizes the tokens in the sequence.
Second, we query all non-buggy examples for those similar to a given buggy example using a nearest neighbor algorithm.
The following explains these two steps in more detail.

The first step converts sequence vectors to frequency vectors.
Let $v_s = [t_1, t_2, \dots, t_n]$ be a vector of $n$ tokens corresponding to code sequence $s$.
We convert $v_s$ into a vector of frequencies $v_{s}^{\mathrm{freq}}$ of all words in $V$.
In other words, we compute:
$$v_s^{\mathrm{freq}} = [count(t_{i_1}, s), count(t_{i_2}, s), \dots, count(t_{i_{|V|}}, s)]$$
for some fixed ordering $i_1, i_2, \dots, i_{|V|}$ of the vocabulary $V$, and where $count(t,s)$ returns the number of occurrences of $t$ in $s$. We exclude the special tokens \code{UNK} and \code{PAD} when computing $v_s^{\mathrm{freq}}$.

Before searching the space of non-buggy examples using the token-frequency vectors, we counteract the effect of tokens with very high frequencies.
Examples of these tokens include \code{new}, \code{=}, \code{return}, and separators, all of which are likely to appear across many different sequences of source code but are less relevant for selecting non-buggy examples.
To counteract their influence, we apply \emph{term frequency-inverse document frequency (TF-IDF)}, which offsets the number of occurrences of each token in the frequency vectors by the number of sequences this token appears in.
TF-IDF is widely used in information retrieval and text mining to reflect how important a word is to a document in a corpus of documents, while accommodating for the fact that some words occur more frequently than others.

As the second step, to search the space of non-buggy code sequences in our data set, we use an efficient, high-dimensional search technique called approximated nearest neighbor (ANN).
We use ANN to search the vector representations of all non-buggy methods for a subset $S_{k_{\mathrm{nBug}}}^{\mathrm{ANN}}$ of non-buggy examples that are similar to the multi-dimensional space of sequence vectors in $S_{k_{\mathrm{bug}}}$.

\begin{definition}[ANN Non-Buggy Examples]
	For every buggy example $s_{k_{\mathrm{bug}}} \in S_{k_{\mathrm{bug}}}$ of bug kind $k \in K$, the ANN of $s_{k_{\mathrm{bug}}}$ is $\mathrm{ANN}_{\mathrm{search}}(s_{k_{\mathrm{bug}}}, S_{k_{\mathrm{nBug}}})$ where $\mathrm{ANN}_{\mathrm{search}}(x,Y)$ returns the ANN of $x$ in $Y$. Therefore, the set of non-buggy nearest neighbors sequences of $S_{k_{\mathrm{bug}}}$ is:
	\begin{equation}
	\begin{split}
		S_{k_{\mathrm{nBug}}}^{\mathrm{ANN}} = \big\{s^\prime \in S_{k_{\mathrm{nBug}}} ~\vert~ & s^\prime = \mathrm{ANN}_{\mathrm{search}}(s, S_{k_{\mathrm{nBug}}})~ \\ 
		&\forall s \in S_{k_{\mathrm{bug}}} \big\}
	\end{split}
	\end{equation}
%	$$ S_{k_{\mathrm{nBug}}}^{\mathrm{ANN}} = \big\{s^\prime \in S_{k_{\mathrm{nBug}}} ~\vert~ s^\prime = \mathrm{ANN}_{\mathrm{search}}(s, S_{k_{\mathrm{nBug}}})~\forall s \in S_{k_{\mathrm{bug}}} \big\} $$
	%	\begin{align*}
	%		S_{k_{nBug}}^{ANN} = \big\{ s^\prime ~\vert~ & s^\prime \in S_{k_{nBug}}~\wedge                                          \\
	%		                                            & s^\prime = ANN_{search}(s, S_{k_{nBug}})~\forall s \in S_{k_{bug}} \big\}
	%	\end{align*}
\end{definition}

ANN uses locality sensitive hashing to perform this high-dimensional space, which is much more efficient than exhaustively computing pair-wise distances between all vectors.

\subsubsection{Selecting Buggy Examples}
When selecting sequences $S_{k_{\mathrm{nBug}}}$ of non-buggy examples, we need to consider whether the location of the bug is within the first $n$ tokens of the method.
A warning $w = (k, l_w, m)$ that flags line $l_{w}$ in method $m$ could fall beyond the sequence $s_m$ extracted from $m$ if the last token of $s_m$, $t_n = (lex, t, l_{t_n})$ has $l_{t_n}  < l_w$.
In other words, it could be that a warning flagged at some method by the oracle occurs at a line beyond the extracted sequence of that method because we limit the sequence length to $n$ tokens.
In such a case, we remove this example from the set of buggy examples of bug kind $k$ and we use it as a non-buggy example.

\subsection{Learning Bug Detection Models}
\label{subsec:rnn}

The remaining step in our neural bug finding approach is training the ML model.
Based on the vector representation of buggy and non-buggy examples of code sequences, we formulate the bug finding problem as binary classification.

\begin{definition}[Bug Finding Problem]
	Given a previously unseen piece of code $C$, the problem $P_k: C \to [0, 1]$ is to predict the probability that $C$ suffers from bug kind $k$, where $0$ means certainly not buggy and $1$ means that $C$ certainly has a bug of kind $k$.
\end{definition}

We train a model to find a bug of kind $k$ in a supervised setup based on two types of training examples: buggy examples ($v_{\mathrm{bug}}$, $1$) and non-buggy examples ($v_{\mathrm{nBug}}$, $0$), where $v_{\mathrm{bug}}$ and $v_{\mathrm{nBug}}$ are the vector representations of buggy and non-buggy code, respectively.
During prediction, we interpret a predicted probability lower than $0.5$ as ``not buggy'', and as ``buggy'' otherwise.

%In principle, we can use any ML classification model to solve the bug prediction problem.
Since we model source code as a sequence of tokens, we employ recurrent neural networks (RNNs) as models.
In particular, we use bi-directional RNN with Long Short Term Memory (LSTM)~\cite{gers1999learning} units.
As the first layer, we have an embedding layer that reduces the one-hot encoding of tokens into a smaller embedding vector of length 50.
For the RNN, we use one hidden bi-directional LSTM layer of size 50.
We apply a dropout of $0.2$ to the hidden layer to avoid overfitting.
The final hidden states of the RNN are fed through a fully connected layer to an output layer of dimension $1$, using the sigmoid activation function.
For the loss function, we choose binary cross entropy, and we train the RNN using the Adam optimizer.
Finally, we use a dynamically calculated batch size based on the size of the training data (10\% of the size of the training set with a maximum of 300).

\subsection{Different Evaluation Settings}
\label{subsec:setup}

We study four different ways of combining training and validation data, summarized in \autoref{table:setups}.
These four ways are combinations of two variants of selecting code examples.
On the one hand, we consider balanced data, i.e., with an equal number of buggy and non-buggy examples.
On the other hand, we consider a stratified split, which maintains a distribution of buggy and non-buggy examples similar to that in all the collected data, allowing us to mimic the frequency of bugs in the real-world.
For instance, assume the total number of samples collected for a specific warning kind is 200 samples, of which 50 (25\%) are buggy and 150 (75\%) are not buggy. If we train the model with 80\% of the data and validate on the remaining 20\%, then a stratified split means the training set has 160 samples, of which 40 (25\%) are buggy and 120 (75\%) are not buggy, and the validation set has 40 samples, of which 10 (25\%) are buggy and 30 (75\%) are not buggy.

Evaluation setups \expBS{} and \expANNBS{} correspond to the scenario of using balanced data for training and stratified split for validation.
In setup \expBS{}, we randomly sample the non-buggy examples to build a balanced training set, while in setup \expANNBS{} we use our novel approximated nearest neighbor (ANN) search for non-buggy examples (Section~\ref{subsec:nonbuggy}).
Since for many of the kinds of warnings the number of collected buggy examples is relatively small for a deep learning task, we additionally evaluate a third setup, \expSS{}, where we utilize all non-buggy data available by doing a stratified split for training and validation.
Finally, setup \expBB{} represents the most traditional setup for binary classifiers, which uses balanced training and balanced validation sets.

\begin{table}
	\centering
	\caption{Setups used to evaluate the neural bug finding models.}
	\label{table:setups}
	\small
	\begin{tabular}{llll}
		\toprule
		Experiment & Training & Validation \\ \midrule
		\expBS{} & Balanced & Stratified \\
		\expANNBS{} & Balanced (ANN sampling) & Stratified \\
		\expSS{} & Stratified & Stratified \\
		\expBB{} & Balanced  & Balanced  \\
		\bottomrule
	\end{tabular}
\end{table}

\section{Implementation}
\label{sec:impl}

We use the JavaParser\footnote{http://javaparser.org/} to parse and tokenize all Java methods in the Qualitas corpus. 
Tokenized methods, warnings generated by Error Prone, their kinds, and locations are stored in JSON files for processing by the models.
Python scikit-learn\footnote{https://scikit-learn.org/} is used to compute the TD-IDF of all examples and NearPy\footnote{http://pixelogik.github.io/NearPy/} is used to find the ANN of each buggy example.
To implement the recurrent neural networks, we build upon Keras and Tensorflow\footnote{https://keras.io/ and https://www.tensorflow.org/}.
\section{Results}
\label{sec:study}

We study the potential of neural bug finding by posing the following research questions:

\begin{itemize}
	\item RQ$_1$: How effective are neural models at identifying common kinds of programming errors?
	\item RQ$_2$: Why does neural bug finding sometimes work?
	\item RQ$_3$: Why does neural bug finding sometimes not work?
	\item RQ$_4$: How does the composition of the training data influence the effectiveness of a neural model?
	\item RQ$_5$: How does the amount of training data influence the effectiveness of a neural model?
	\item RQ$_6$: What pitfalls exist when evaluating neural bug finding?
\end{itemize}

\subsection{Experimental Setup}

For each experiment, we split all available data into 80\% training data and 20\% validation data, and we report the results with the validation set.
Each experiment is repeated five times, and we report the average results.
For the qualitative parts of our study, we systematically inspected at least ten, often many more, validation samples from each warning kind.
All experiments are performed on a machine with 48
Intel Xeon E5-2650 CPU cores, 64GB of memory, and an NVIDIA Tesla P100 GPU.

\begin{table*}
	\centering
	\caption{Precision, recall, and f1 of the neural bug finding models of the top \topWarnings{} warnings reported by Error Prone. Results are obtained by training with 80\% of available data and validating on the remaining 20\%. Table also shows the total number of examples available in the data set. Warnings are in descending order by their total number of buggy examples.}
	\label{table:results}
	\begin{adjustbox}{center,width=0.91\textwidth}
	\begin{tabular}{@{}rl@{}rr|rrr|rrr|rrr|rrr@{}}
		\toprule
		& & & & \multicolumn{3}{c|}{\textbf{Experiment~\expBS{}}} & \multicolumn{3}{c|}{\textbf{Experiment~\expANNBS{}}} & \multicolumn{3}{c|}{\textbf{Experiment~\expSS{}}} & \multicolumn{3}{c}{\textbf{Experiment~\expBB{}}} \\ \cmidrule{3-16}
%		& & & \textbf{Training} & \multicolumn{3}{l|}{Balanced} & \multicolumn{3}{l|}{ANN Balanced} & \multicolumn{3}{l|}{Stratified} & \multicolumn{3}{l}{Balanced} \\
%		& & & \textbf{Validation} & \multicolumn{3}{l|}{Stratified} & \multicolumn{3}{l|}{Stratified} & \multicolumn{3}{l|}{Stratified} & \multicolumn{3}{l}{Balanced} \\ \cmidrule{3-16}
		& \multicolumn{3}{r|}{\small{Nb. of examples}} & Pr. & Re. & F1 & Pr. & Re. & F1 & Pr. & Re. & F1  & Pr. & Re. & F1 \\
		Id & Warning & \small{Buggy} & \small{nBuggy} & \% & \% & \% & \% & \% & \% & \% & \% & \%  & \% & \% & \% \\ \midrule
		1 & MissingOverride~\textsuperscript{G} & 268,304 & 501,937 & 69.74 & 86.05 & 76.97 & 73.53 & 77.70 & 75.48 & 79.78 & 74.97 & 77.28 & 82.34 & 84.25 & 83.24 \\
		2 & BoxedPrimitiveConstructor~\textsuperscript{L} & 3,769 & 767,112 & 12.00 & 96.47 & 21.23 & 17.47 & 93.93 & 29.20 & 93.62 & 92.02 & 92.67 & 95.51 & 94.26 & 94.85 \\
		3 & Sync.OnNonFinalField~\textsuperscript{G} & 2,282 & 653,856 & 20.19 & 98.73 & 33.18 & 24.14 & 97.76 & 38.57 & 71.05 & 79.43 & 74.88 & 96.28 & 99.74 & 97.97 \\
		4 & ReferenceEquality~\textsuperscript{G} & 1,680 & 746,285 & 1.48 & 89.17 & 2.90 & 1.55 & 83.21 & 3.05 & 78.94 & 39.40 & 52.08 & 85.01 & 90.40 & 87.51 \\
		5 & DefaultCharset~\textsuperscript{G} & 1,550 & 747,192 & 2.18 & 95.35 & 4.27 & 4.06 & 80.00 & 7.69 & 75.61 & 60.58 & 66.57 & 91.83 & 94.56 & 93.13 \\
		6 & EqualsHashCode~\textsuperscript{G} & 590 & 673,446 & 8.20 & 99.49 & 14.91 & 8.79 & 85.25 & 15.89 & 39.71 & 5.93 & 10.06 & 98.38 & 100.00 & 99.17 \\
		7 & Unsync.OverridesSync.~\textsuperscript{G} & 517 & 657,303 & 0.36 & 82.14 & 0.72 & 0.28 & 68.93 & 0.55 & 61.26 & 16.89 & 25.27 & 85.74 & 77.05 & 80.73 \\
		8 & ClassNewInstance~\textsuperscript{G} & 486 & 742,585 & 0.80 & 94.23 & 1.59 & 2.56 & 85.36 & 4.97 & 88.04 & 79.59 & 83.46 & 91.41 & 93.97 & 92.44 \\
		9 & OperatorPrecedence~\textsuperscript{L} & 362 & 716,691 & 0.51 & 92.22 & 1.02 & 0.49 & 75.56 & 0.98 & 70.10 & 20.28 & 30.00 & 89.91 & 88.67 & 89.17 \\
		10 & DoubleCheckedLocking~\textsuperscript{G} & 204 & 297,959 & 2.80 & 97.56 & 5.40 & 5.05 & 95.61 & 9.24 & 95.80 & 83.41 & 88.84 & 98.30 & 95.53 & 96.77 \\
		11 & NonOverridingEquals~\textsuperscript{L} & 165 & 488,094 & 2.04 & 93.33 & 3.94 & 2.97 & 77.58 & 5.61 & 90.01 & 87.88 & 88.63 & 95.22 & 97.95 & 96.49 \\
		12 & NarrowingCompoundAssign.~\textsuperscript{L} & 158 & 660,390 & 0.29 & 88.12 & 0.58 & 0.34 & 79.38 & 0.68 & 53.04 & 31.25 & 38.11 & 92.72 & 92.22 & 92.45 \\
		13 & ShortCircuitBoolean~\textsuperscript{L} & 116 & 616,037 & 0.09 & 82.61 & 0.18 & 0.10 & 73.91 & 0.20 & 72.22 & 31.30 & 39.70 & 78.21 & 91.82 & 83.78 \\
		14 & IntLongMath~\textsuperscript{L} & 111 & 531,502 & 0.23 & 79.09 & 0.47 & 0.30 & 81.82 & 0.59 & 59.52 & 7.27 & 12.60 & 90.82 & 100.00 & 94.95 \\
		15 & NonAtomicVolatileUpdate~\textsuperscript{G} & 80 & 369,501 & 0.07 & 71.25 & 0.15 & 0.04 & 71.25 & 0.08 & 0.00 & 0.00 & 0.00 & 80.24 & 83.60 & 81.00 \\
		16 & WaitNotInLoop~\textsuperscript{G} & 77 & 469,210 & 0.27 & 97.33 & 0.53 & 0.30 & 86.67 & 0.59 & 83.17 & 49.33 & 61.52 & 89.75 & 100.00 & 94.57 \\
		17 & ArrayToString~\textsuperscript{L} & 56 & 554,213 & 0.07 & 96.36 & 0.13 & 0.04 & 61.82 & 0.08 & 20.00 & 1.82 & 3.33 & 96.36 & 96.67 & 96.18 \\
		18 & MissingCasesInEnumSwitch~\textsuperscript{G} & 53 & 430,701 & 0.10 & 85.45 & 0.20 & 0.05 & 43.64 & 0.10 & 0.00 & 0.00 & 0.00 & 81.97 & 94.64 & 87.09 \\
		19 & TypeParam.UnusedInFormals~\textsuperscript{L} & 46 & 321,451 & 0.41 & 86.67 & 0.81 & 0.69 & 93.33 & 1.35 & 0.00 & 0.00 & 0.00 & 92.70 & 93.33 & 92.50 \\
		20 & FallThrough~\textsuperscript{L} & 45 & 615,140 & 0.08 & 93.33 & 0.15 & 0.43 & 82.22 & 0.84 & 63.33 & 20.00 & 30.09 & 83.44 & 92.29 & 87.13 \\ \midrule
		   & \textbf{Median} & & & 0.46 & 92.78 & 0.92 & 0.59 & 80.91 & 1.17 & 70.58 & 31.28 & 38.91 & 91.12 & 94.12 & 92.48 \\ 
		\bottomrule
	\end{tabular}
	\end{adjustbox}
\end{table*}

%% RQ1 %%
\subsection{RQ$_1$: How effective are neural models at identifying common kinds of programming errors?}
\label{subsec:rq1}

To study the effectiveness of the neural bug finding models, we measure their precision, recall, and F1-score.
For a specific bug kind, precision is the percentage of actual bugs among all methods that the model flags as buggy, and recall is the percentage of bugs detected by the model among all actual bugs.
The F1-score is the harmonic mean of precision and recall.

We first look at Experiment~\expANNBS{}, which uses balanced training data selected using ANN and an imbalanced validation set.
The results of this and the other experiments are shown in~\autoref{table:results}.
Across the \topWarnings{} kinds of warnings we study, precision ranges between 73.5\% down to 0.04\%, while recall ranges between 97.76\% and 43.6\%. 
The relatively high recall shows that neural bug finders find a surprisingly high fraction of all bugs.
However, as indicated by the low precision for many warnings kinds, many of the models tend to report many spurious warnings.

In Experiment~\expSS{}, we use a much larger, but imbalanced, training set.
\autoref{table:results} also shows the results of this experiment.
One can observe a clear improvement of precision over Experiment~\expANNBS{} for many of the models.
This improvement in precision is due to the richer and larger training set, which trains the model with many more non-buggy examples than Experiment~\expANNBS{}, making it more robust against false positives. 
However, the increased precision comes at the cost of decreasing recall compared to Experiment~\expANNBS{}.
For example, the neural model that predicts double checked locking bugs (Id~10 in~\autoref{table:results}) has its recall dropping from 95.6\% to 83.4\% when using the full training data available. 
Yet, the reduced recall is offset by a huge increase in precision, causing the median F1-score to grow from 1.17\% in Experiment~\expANNBS{} to 38.91\% in Experiment~\expSS{}.

\begin{resultbox}
The effectiveness of neural bug finders varies heavily across bug patterns, reaching a precision of up to 95.8\% and a recall of up to 97.76\% for some patterns, while dropping down to almost 0\% for others. 
%Our neural models find bugs with a precision up to 95.8\% and a recall up to 97.76\% depending on the kind of bug. For a few of the warning kinds, precision, recall, or both drop down to almost zero because the model is unable to learn the underlying pattern.
\end{resultbox}

%% RQ2 %%
\subsection{RQ$_2$: Why does neural bug finding work?}
To answer this question and also RQ$_3$, we systematically inspect true positives, true negatives, false positives, and false negatives for each model.
We discuss our observations by splitting the warning kinds into two groups, based on whether the information provided to the neural model is, in principle, sufficient to accurately detect the kind of bug.

\subsubsection{Bug Kinds with Sufficient Available Information}

The first group includes all bug kinds where the bug pattern could, in principle, be precisely and soundly detected based on the information we provide to the neural model.
Recall that we feed the first 50 tokens of a method into the model, and no other information, such as the class hierarchy or other methods in the program.
In other words, the model is given enough information to reason about local bugs, which involve a property of one or a few statements, one or a few expressions, or the method signature.
We mark all warning kinds in this group with a~\textsuperscript{L} (for local) in \autoref{table:results}.
Intuitively, these warning kinds correspond to what traditional lint-like tools may detect based on a local static analysis.

We now discuss examples of true positives, i.e., correctly identified bugs, among the warnings reported by models trained for warning kinds in the first group.

\paragraph{Boxed primitive constructor (Id~2)}
This bug pattern includes any use of constructors of primitive wrappers, such as \code{new Integer(1)} and \code{new Boolean(false)}. 
The neural bug finder for this warning achieves high precision and recall of 93.6\% and 92\% respectively (\autoref{table:results}, Experiment~\expSS{}).
The following is an instance of this bug, which is detected by the neural model:
%{
%	"prediction": "1.0",
%	"comment": "",
%	"path:": "/home/ah/codeAugmentation/Data/qualitas/QualitasCorpus-20130901r/Systems/sandmark/sandmark-3.4/src/smark3/sandmark/util/graph/graphview/NodeDisplayInfo.java",
%	"line": 50
%}
\begin{lstlisting}
public int compareTo(java.lang.Object o) {
 return new Integer(myX).compareTo(new Integer(((NodeDisplayInfo)o).myX));
}
\end{lstlisting}
Inspecting these and other bug kinds shows that, in essence, the model learns to identify specific subsequences of tokens, such as "\dots~\code{new Boolean}~\dots" and "\dots~\code{new Integer}~\dots", as a strong signal for a bug.

\paragraph{Operator precedence (Id~9)}
This warning is about binary expressions that either involve ungrouped conditionals, such as \code{x || y \&\& z}, or a combination of bit operators and arithmetic operators, such as \code{x + y <{}< 2}.
Such expressions are confusing to many developers and should be avoided or made more clear by adding parentheses.
The following is a true positive detected by our neural model.
%{
%"comment": "",
%"path:": "/home/ah/codeAugmentation/Data/qualitas/QualitasCorpus-20130901r/Systems/lucene/lucene-4.3.0/src/lucene-4.3.0/codecs/src/java/org/apache/lucene/codecs/simpletext/SimpleTextTermVectorsReader.java", 
%"line": 506,
%"prediction": "0.6"
%}
\begin{lstlisting}
@Override
public int nextPosition() {
 assert (positions != null && 
 	nextPos < positions.length) 
 	|| startOffsets != null 
 	&& nextPos < startOffsets.length;
 '\dots'
}
\end{lstlisting}
Overall, the neural model achieves 70\% precision and 20.28\% recall.
The fact that the model is relatively successful shows that neural bug finders can learn to spot non-trivial syntactic patterns.
Note that the space of buggy code examples for this warning kind is large,
because developers may combine an arbitrary number of binary operators and operands in a single statement.
Given that the model is trained on very few buggy examples, 290 (80\% of 362), the achieved precision and recall are promising. 

\begin{resultbox}
The models learn syntactic patterns commonly correlated with particular kinds of bugs and identify specific tokens and token sequences, such as calls to particular APIs.
\end{resultbox}

\subsubsection{Bug Kinds with Only Partial Information}

The second group of bug kinds contains bug patterns that, in principle, require more information than available in the token sequences we give to the neural models to be detected soundly and precisely.
For example, detecting these kinds of bugs requires information about the class hierarchy or whether a field used in a method is final.
We mark these bug kinds with a~\textsuperscript{G} (for global) in \autoref{table:results}.
The bug kinds include bugs that require type and inheritance information, e.g., missing override annotations (Id~1), missing cases in enum switch (Id~18), default Charset (Id~5), and un-synchronized method overriding a synchronized method (Id~7).
They also include bugs for which some important information is available only outside the current method, such as synchronized on non-final field (Id~3) and equals-hashcode (Id~6).
Note that although detecting these bugs requires information beyond the sequence of tokens extracted from the methods, the bug location lies within the sequence of tokens.
Somewhat surprisingly, neural bug finding also works for some of these bug patterns, achieving precision and recall above 70\% in some cases, which we describe in the following.

\paragraph{Missing \code{@Override} (Id~1)}
This warning is for methods that override a method of an ancestor class but that do not annotate the overriding methods with \code{@Override}.
Although the supertype information that is required to accurately detect this problem is not available to the neural model, the model provides high precision and recall.
Inspecting true positives and training examples reveals that the model learns that many overriding methods override methods of common Java interfaces and base classes. 
Examples include the \code{toString()} method from the \code{Object} base class and the \code{run()} method from the \code{Runnable} interface.
In fact, both method names appear in the data set as buggy 44,789 and 21,767 times, respectively.
In other words, the models successfully learns to identify common instances of the bug pattern, without fully learning the underlying bug pattern.

\paragraph{Default \code{Charset} (Id~5)}
This warning flags specific API usages that rely on the default Charset of the Java VM, which is discouraged for lack of portability.
The ``pattern'' to learn here are specific API names, which implicitly use the default Charset.
The following instance is a true positive detected by the neural model:
%{
%	"path:": "/home/ah/codeAugmentation/Data/qualitas/QualitasCorpus-20130901r/Systems/findbugs/findbugs-1.3.9/src/findbugs-1.3.9/src/java/edu/umd/cs/findbugs/detect/NoteCheckReturnValue.java",
%	"prediction": "0.98",
%	"line": 124,
%	"comment": ""
%}
\begin{lstlisting}
private void saveTraining() {
  BufferedWriter writer = null;
  try {
    writer = new BufferedWriter(new FileWriter(SAVE_TRAINING));
    '\dots'
\end{lstlisting}
As we show in RQ$_3$, this bug is more subtle than it looks.
Correctly detecting this problem requires, in some cases, information on the type of receiver objects, on which the APIs are called.

\paragraph{Double checked locking (Id~10)}
This bug is about a lazy initialization pattern~\cite{lazyInit} where an object is checked twice for nullness with synchronization in-between the null checks, to prevent other threads from initializing the object concurrently.
The following is a true positive reported by our neural model.\footnote{Note that our approach extracts the token sequence from the method body, i.e., starting from line 3. The object declaration at line 1 is shown for completeness only.}
%{
%	"line": 33,
%	"path:": "/home/ah/codeAugmentation/Data/qualitas/QualitasCorpus-20130901r/Systems/aspectj/aspectj-1.6.9/src/org/aspectj/org/eclipse/jdt/core/dom/ReferencePointcut.java",
%	"prediction": "0.98",
%	"comment": ""
%}
\begin{lstlisting}
private SimpleName pointcutName = null;
'\dots'
public SimpleName getName() {
  if (this.pointcutName == null) {
    synchronized (this) {
      if (this.pointcutName == null) {
        '\dots'
  return this.pointcutName;
}
\end{lstlisting}
While the method with the bug contains parts of the evidence for the bug, it is missing the fact that the field \code{pointcutName} is not declared as \code{volatile}.
So how does the model for this bug pattern achieve the surprisingly high precision and recall of 95.8\% and  83.41\%, respectively (Experiment \expSS{})?
We find that the correct pattern of double checked locking almost never occurs in the data set.
Even the ANN search for non-buggy examples yields sequences that are indeed similar, e.g., sequences that have a null check followed by a synchronized block, but that do not exactly match the lazy initialization pattern. 
Given the data set, the model learns that a null check, followed by a synchronized block, followed by a another null check is likely to be buggy.
In practice, this reasoning seems mostly accurate, because the idiom of double checked locking is hard to get right even for experienced programmers~\cite{doubleCheckedLock}.

\begin{resultbox}
	Neural bug finding sometimes works even when only parts of the information to accurately detect a specific kind of bug is given.
	The reason is that models learn to identify common instances of the general bug pattern or simply ignore unlikely side conditions.
\end{resultbox}

%% RQ3 %%
\subsection{RQ$_3$: Why does neural bug finding sometimes not work?}
To answer this question, we systematically inspect false positives and false negatives for each model.
We present one example for each case and provide insights why the models mis-classify them.

\subsubsection{Spurious Warnings}
Spurious warnings, i.e. false positives, occur when a model predicts a non-existing bug.

\paragraph{Default Charset (Id~5)}
In RQ$_2$, we showed that finding this bug pattern entails learning specific API names, e.g., \code{FileWriter}.
Another common API that raises this warning is \code{String.getBytes()}, which also relies on the platform default Charset. 
Because this API is strongly present in the training examples, the model learns that sequences that have the \code{getBytes} token are likely to be buggy.
However, whether an occurence of this token is erroneous depends on the receiver object on which the method is called.
The following is a false positive for this bug kind, where a method with the same name is declared for a user defined type.
%{
%	"path:": "/home/ah/codeAugmentation/Data/qualitas/QualitasCorpus-20130901r/Systems/aspectj/aspectj-1.6.9/src/org/aspectj/weaver/bcel/UnwovenClassFile.java",
%	"prediction": "0.66",
%	"line": 55,
%	"comment": ""
%}
\begin{lstlisting}
public class UnwovenClassFile implements IUnwovenClassFile {
  '\dots'
  public byte[] getBytes() {
    return bytes;
  }
  '\dots'
\end{lstlisting}

\subsubsection{Missed Bugs}
The neural models inevitably have false negatives, i.e., they fail to detect some instances of the bug patterns.

\paragraph{Non-overriding equals (Id~11)}
This bug pattern flags methods which look like \code{Object.equals}, but are in fact different.
A method overriding \code{Object.equals} must have the parameter passed to it strictly of type \code{Object}, a requirement for proper overload resolution.
Therefore, any method that looks like \code{boolean equals(NotObjectType foo) \{\dots\}} should be flagged buggy.
The following, is an instance of a false negative for this warning kind.
%{
%	"prediction": "0.21",
%	"line": 493,
%	"comment": "",
%	"path:": "/home/ah/codeAugmentation/Data/qualitas/QualitasCorpus-20130901r/Systems/hsqldb/hsqldb-2.0.0/src/hsqldb-2.0.0/hsqldb/src/org/hsqldb/index/NodeAVLDisk.java"
%}
\begin{lstlisting}
boolean equals(NodeAVL n) {
 if (n instanceof NodeAVLDisk) {
  return this == n || 
  (getPos() == ((NodeAVLDisk) n).getPos());
 }
 return false;
}
\end{lstlisting}
The reason why the model misses this bug is that it fails to distinguish between ``\code{boolean equals(Object}'' and any other sequence ``\code{boolean equals(NotObjectType}''.
We believe that this failure is not an inherent limitation of the neural model, but can rather be attributed to the scarcity of our training data.
In total, we have 165 examples of this bug in our data set, and for training the model, we use 80\% of the data, i.e., around 132 examples.
Given this amount of data, the recall for this bug reaches 87.88\% (Experiment~\expSS{}).

%% RQ4 %%
\subsection{RQ$_4$: How does the composition of the training data influence the effectiveness of a neural model?}
\label{subsec:rq4}
To answer this question, we compare the results from Experiments~\expBS{},~\expANNBS{}, and~\expSS{}.
Comparing Experiments~\expBS{} and~\expANNBS{} in~\autoref{table:results} shows that using ANN to select non-buggy samples for training increases the precision of the trained models in most of the cases.
The reason is that having similar code examples, some of which are labeled as buggy while others are labeled non-buggy, helps the model to define a more accurate border between the two classes.
Recent work on selecting inputs for testing neural networks is based on a similar observation~\cite{Kim2018}.
At the same time, using ANN also causes a drop in recall, mainly because the model faces a more difficult learning task.
For example, using ANN to train the model for bug pattern~2 improves precision by 5.5\% but degrades recall by 2.5\%.

Comparing Experiments~\expANNBS{} and~\expSS{} shows that adding more non-buggy examples to the training set decreases the recall by a value between 2\% (bug pattern~2) up to a complete erasure of the recall (bug pattern~19). 
On the positive side, the additional data added in Experiment~\expSS{} significantly improves the precision of all models.
For example, the model of bug pattern~16 improves precision by 83\%.

\begin{resultbox}
The composition of the training data has a huge impact.
Balanced training data (Experiments~\expBS{} and~\expANNBS{}) favors recall over precision, while adding more non-buggy training data (Experiment~\expSS{}) favors precision.
\end{resultbox}

%% RQ5 %%
\subsection{RQ$_5$: How does the amount of training data influence the effectiveness of a neural model?}

\begin{figure*}\centering
	~\hspace{-1.9em}
	\subfigure[Experiment~\expANNBS{} precision]{\label{}\includegraphics[width=.26\linewidth]{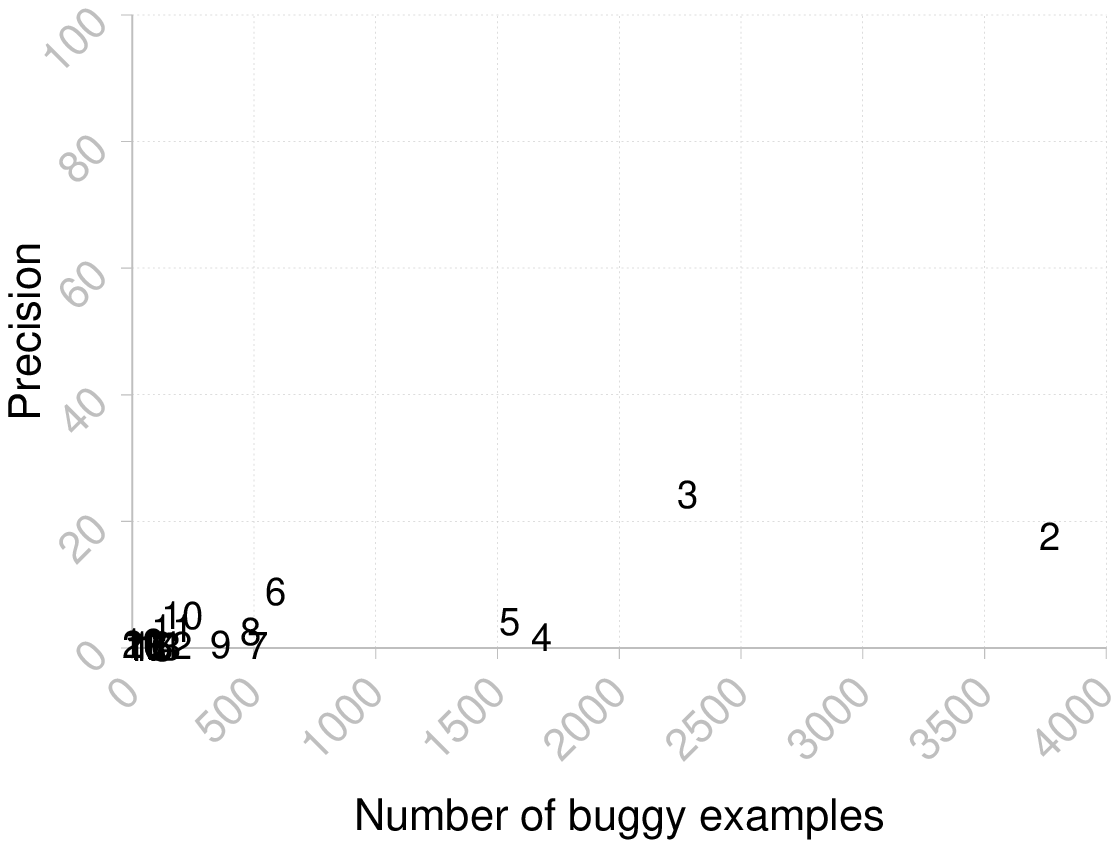}}
	~\hspace{-1.9em}
	\subfigure[Experiment~\expANNBS{} recall]{\label{}\includegraphics[width=.26\linewidth]{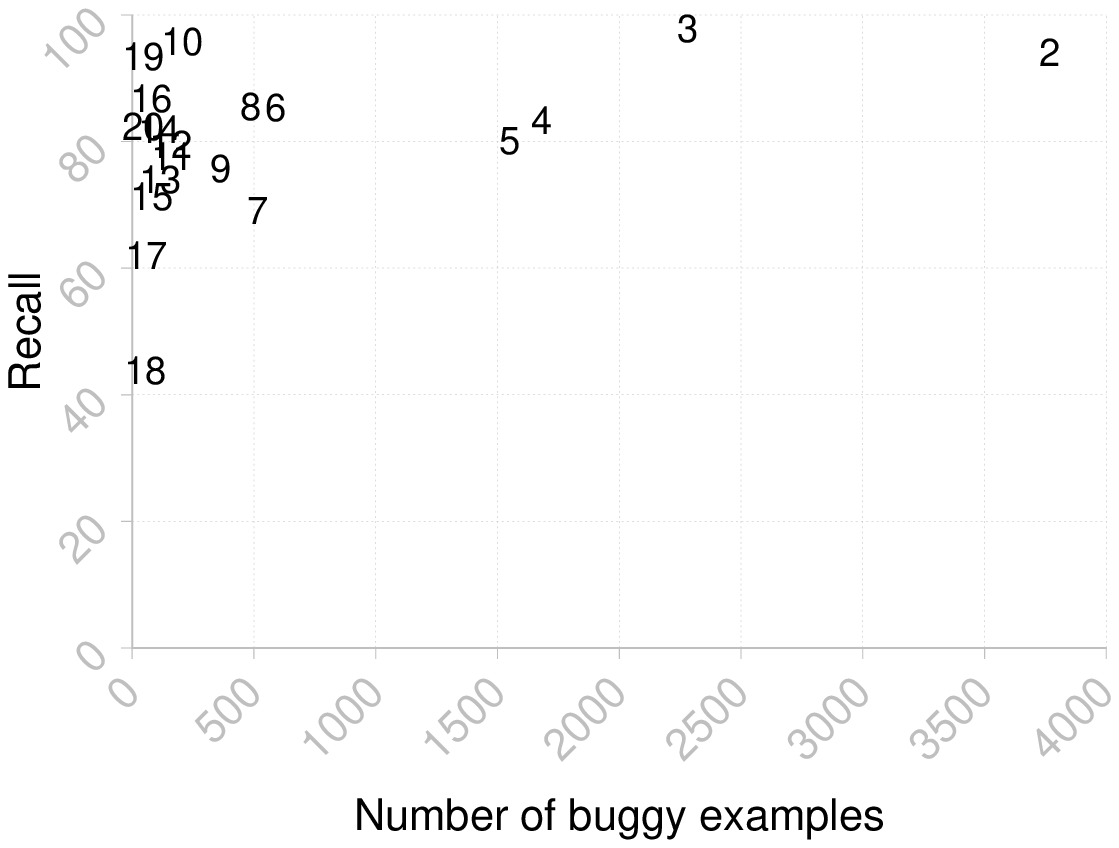}}
	~\hspace{-1.9em}
	\subfigure[Experiment~\expSS{} precision]{\label{}\includegraphics[width=.26\linewidth]{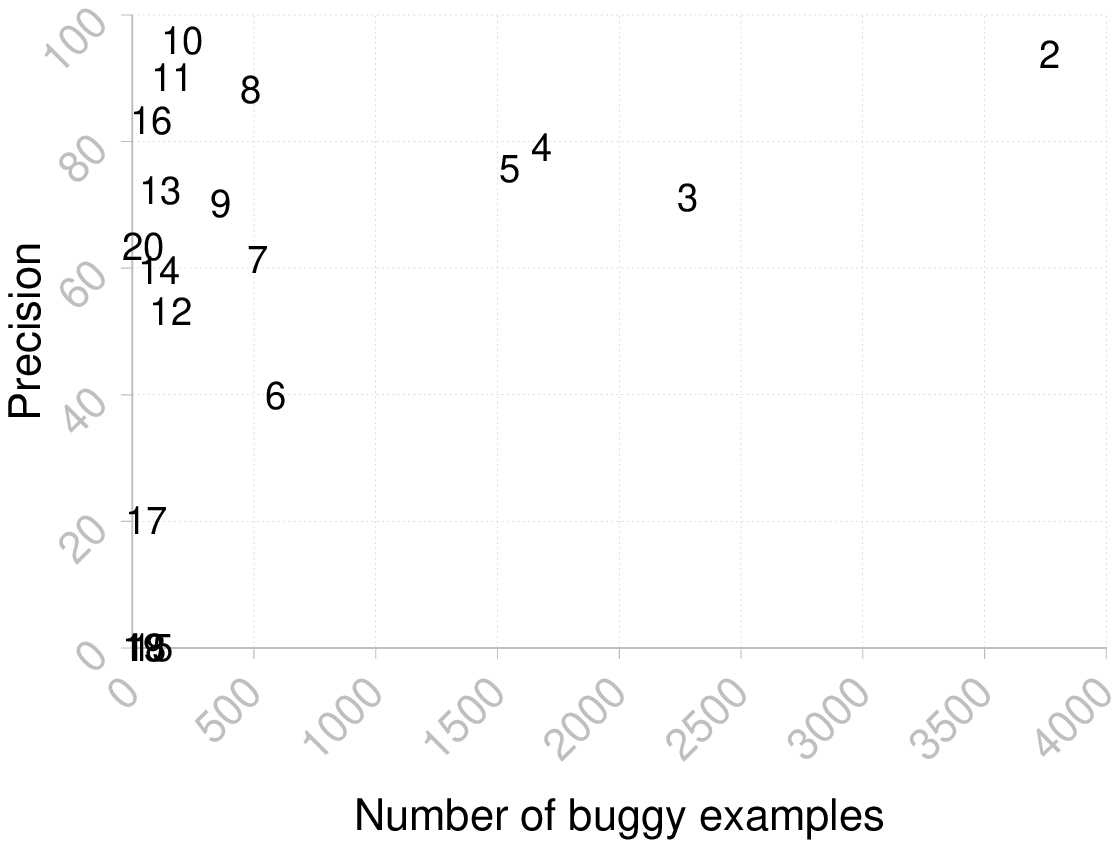}}
	~\hspace{-1.9em}
	\subfigure[Experiment~\expSS{} recall]{\label{}\includegraphics[width=.26\linewidth]{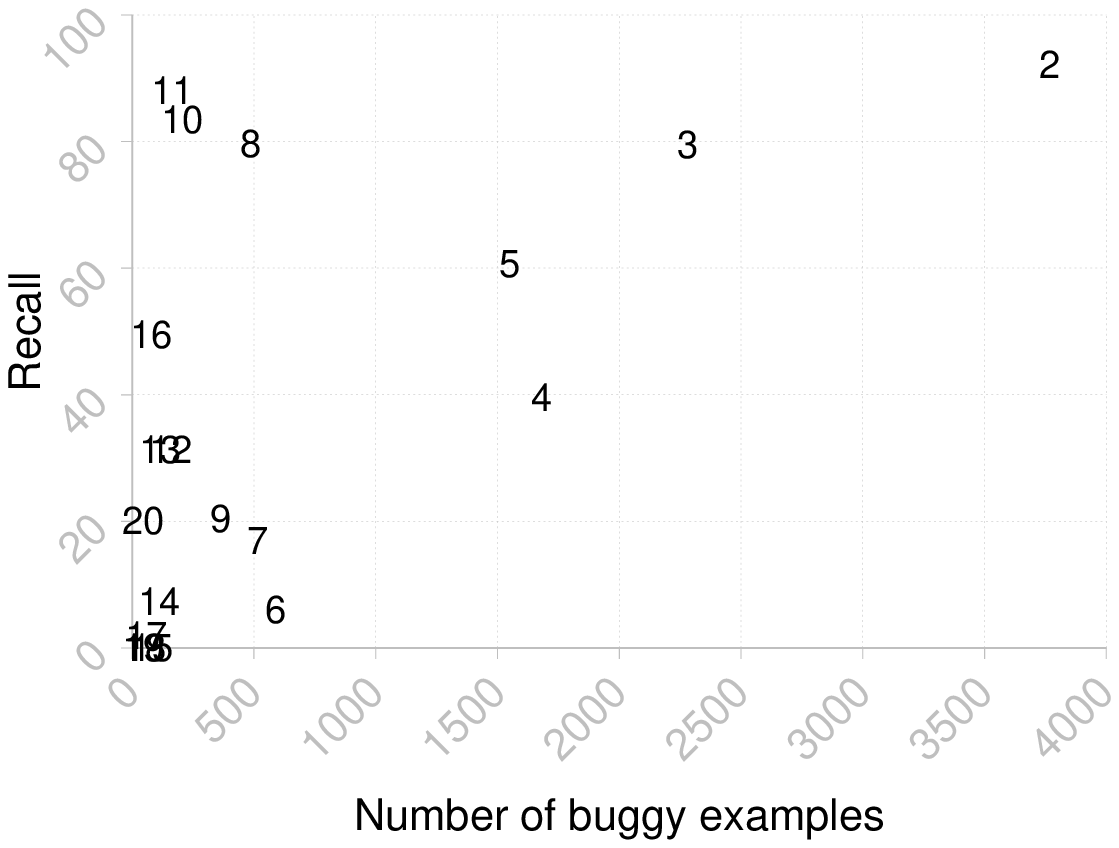}}
	
	\caption{Effect of number of buggy examples on precision and recall for each warning kind. The plots use the ids from~\autoref{table:results}. Bug Id~1 is not shown due to the huge difference in x-axis scale.}
	\label{fig:scatter}
\end{figure*}

\autoref{fig:scatter} addresses this question by plotting precision and recall of the different models over the number of buggy examples that a model is trained on.
All four plots show a generally increasing effectiveness, both in terms of precision and recall, for warning kinds, where more data is available.
For example, the models for bug patterns~2 and~3 reach high precision and recall in both experiments~\expANNBS{} and~\expSS{} due to the availability of more examples.
Perhaps surprisingly, though, some models are effective even with much a smaller number of warnings.
For example, for bug patterns~11 and~16, the neural models achieve precision and recall above 77\%, even though only 165 and 77 buggy examples are available, respectively.

\begin{resultbox}
More training data improves the effectiveness of a learned model, but surprisingly small data sets, e.g., of only 77 buggy examples, can yield effective models.
\end{resultbox}

%% RQ6 %%
\subsection{RQ$_6$: What pitfalls exist when evaluating neural bug finding?}
\label{subsec:rq6}

In binary classification problems, the usual setup for training and validation is to use balanced data sets.
However, bugs of a specific kind are rare in real-world code. Therefore, evaluating neural bug finding\footnote{and any bug finding technique} using a balanced data setup yields misleading results, as described in the following.

\autoref{table:results} shows the results of Experiment~\expBB{}, which uses balanced data for both training and validation.
The first glimpse at the results is very encouraging, as they show that neural bug finding works pretty well.
Unfortunately, these numbers are misleading.
The reason for the spuriously good results is that the neural models overfit to the presence, or absence, of particular tokens, which may not necessarily be strong indicators of a bug.

As an example, consider bug pattern~6 , which flags classes that override the \code{Object.equals} method but that fail to also override \code{Object.hashCode}.
In~\autoref{table:results}, Experiment~\expBB{}, the neural model predicting this warning is almost perfect with 100\% recall and 98.38\% precision.
However, a closer look into this model and manual inspection of the training and validation examples reveal that the neural model has simply learned to predict that the sequence of tokens "\code{public boolean equals (Object}~\dots" is always buggy.
This explains why the model achieves a recall of 100\%.
But why is precision also quite high at 98\%?
It turns out that randomly sampling 590 non-buggy examples (corresponding to the number of buggy examples) from 673,446 non-buggy methods is likely to yield mostly methods that do not contain the sequence "\code{public boolean equal}~\dots".
In other words, the unrealistic setup of training and validation data misleads the model into an over-simplified task, and hence the spuriously good results.

Comparing the results from Experiments~\expBS{} and~\expBB{} further reveals the fragility of Experiment~\expBB{}'s setup.
In Experiment~\expBS{}, the training set is constructed as in Experiment~\expBB{}, but the validation set contains a lot more samples, most of them are actually not buggy.
Because the models learned in Experiment~\expBB{} do not learn to handle non-buggy examples similar to the buggy examples, their precision is low.
That is why for the same warning kind, e.g. Id~6, the precision in Experiment~\expBS{} is only 8\% instead of the 98.38\% in Experiment~\expBB{}.

\begin{resultbox}
Even though bug detection can be seen as binary classification tasks, evaluating its effectiveness with balanced validation data can be highly misleading.
\end{resultbox}

\subsection{Data Availability}

All data required to inspect and reproduce our results will be made publicly available with the final version of the paper.

\section{Discussion}
\label{sec:reflection}

\subsection{Lessons Learned}
\label{subsec:xxx}

The overall question of this paper is whether neural bug finding is feasible.
Given our results, we give a positive yet cautious answer.
We see empirically that neural models can learn syntactic code patterns, and hence these models are indeed capable of finding local bugs that do not require inter-procedural or type-based reasoning.
Moreover, even for the more difficult bugs, which require information beyond the sequence of tokens extracted from methods, e.g. type and inter-procedural information, simple sequence-based learning surprisingly detects a non-negligible percentage of the bugs.

To make neural bug finding applicable to wider range of bugs, our work reveals the need for richer ML models that utilize information beyond the source code tokens, e.g., type hierarchy, API-specific knowledge, and inter-procedural analysis.
How to effectively feed such information into neural models is closely related to the ongoing challenge of finding suitable source code representations for machine learning.

Finally, our results emphasize another long-standing challenge in machine learning: data is important.
Our results demonstrate that both the amount of training data as well as how to sample the training data has a huge influence on the effectiveness of the learned bug finding models.
Collecting data for neural bug finding remains an open problem, which seems worthwhile addressing in future work.

%\begin{itemize}
%	\item RNN is capable of detecting local syntactic bugs
%	\item Even for global bugs, simple sequence-based patterns detect a large number of these bugs

%	\item Richer ML models for source code with more information e.g. type hierarchy
%	\item Data is important: research on generating/collecting more data
%\end{itemize}

\subsection{Threats to Validity}
\label{subsec:threats}
Our training and validation subjects might bias the results towards these specific projects, and the findings may not generalize beyond them.
We try to mitigate this problem by using the Qualitas corpus, which consists of a diverse set of 112 real-world projects.

We use warnings reported by a static analyzer as a proxy for bugs.
The fact that some of these warnings may be false positives and that some actual bugs may be missed, creates some degree of noise in our ground truth.
By building upon an industrially used static analyzer tuned to have less than 10\% false positives~\cite{sadowski2015tricorder}, we try to keep this noise within reasonable bounds.
Future research on collecting and generating buggy and non-buggy code examples will further mitigate this problem.

Finally, the qualitative analysis of the validation results is subject to human error. 
To mitigate this, two of the authors discussed and validated all the findings.
\section{Related Work}
\label{sec:related}

\subsection{Static Bug Finding}

Techniques for scanning source code for particular bug patterns go back to the pioneering lint tool~\cite{Johnson1978}.
More recent tools deployed in industry include Error Prone~\cite{Aftandilian2012}, which is used at Google and serves as an oracle for our study, and Infer~\cite{Calcagno2015}, which is used at Facebook.
Detailed accounts of deploying static bug detectors consider a name-based static checker~\cite{oopsla2017}, applying the FindBugs tool~\cite{Ayewah2008,Ayewah2010}, and a rule inference-based static bug detector~\cite{Bessey2010}.
Research on static bug finding includes work on finding API
misuses~\cite{Wasylkowski2009,Nguyen2009,icse2012-statically},
name-based bug detection~\cite{issta2011},
security bugs~\cite{Brown2017},
finding violations of inferred programmer beliefs~\cite{Engler2001}, and
other kinds of anomaly detection~\cite{Liang2016}.
These approaches involve significant manual effort for creating and tuning the bug detectors, whereas we here study bug detectors learned from examples only.

The presence of false positives, a problem shared by both traditional and learned bug detectors, motivates work on prioritizing analysis warnings, e.g., based on the frequency of true and false positives~\cite{Kremenek2003},
the version history of a program~\cite{Kim2007},
and statistical models based on features of warnings and
code~\cite{Ruthruff2008}.
These efforts are orthogonal to the bug detection problem addressed in this paper, and could possibly be combined with neural bug detectors.

\subsection{Machine Learning and Language Modeling for Bug Finding}	

% classification
Learned models are becoming increasingly popular for bug finding.
DeepBugs exploits identifier names, e.g., of variables and methods, to find buggy code~\cite{oopsla2018-DeepBugs}.
Vasic et al.~\cite{Vasic2019} use pointer networks to jointly find and fix variable mis-use bugs.
Choi et al.\ train a memory network~\cite{Weston2014} to predict whether a piece of code may cause a buffer overrun~\cite{Choi2017}.
A broader set of coding mistakes that may cause vulnerabilities is considered in other learning-based work~\cite{Li2018a}.
Harer et at.~\cite{Harer2018} train a CNN to classify methods as vulnerable or not based on heuristics built on labels from a static analyzer.
%These approaches and our work share the idea of formulating bug detection as a classification problem.
The main contribution of our work is to systematically study general neural bug detection and to predict the bug kind.

% anomaly detection
Instead of classifying whether a piece of code suffers from a bug, anomaly detection approaches search for code that stands out and therefore may be buggy.
Bugram uses a statistical language model that warns about uncommon n-grams of tokens~\cite{Wang2016}.
Salento learns a probabilistic model of API usages and warns about unusual usages~\cite{Murali2017}.
In contrast to our work, these techniques learn from non-buggy examples only.
Ray et al.~\cite{Ray2016} explains why this is possible and shows that buggy code is less natural than non-buggy code.

Orthogonal to bug detection is the problem of defect prediction~\cite{fenton1999critique,zimmermann2009cross}.
Instead of pinpointing specific kinds of errors, as our work, it predicts whether a given software component will suffer from any bug at all.
Wang et al.~\cite{Wang2016a} propose a neural network-based model for this task~\cite{Wang2016a}.

\subsection{Machine Learning on Programs}

Beyond bug detection, machine learning has been applied to other programming-related tasks~\cite{Allamanis2018}, such as 
predicting identifier names~\cite{Raychev2015,Context2Name} and
types~\cite{Raychev2015,Hellendoorn2018,icse2019}.
A challenge for any learning-based program analysis is how to represent code.
Work on this problem includes graph-based representations~\cite{Allamanis2017b,Brockschmidt2018},
embeddings learned from sequences of API calls~\cite{DeFreez2018},
embeddings learned from paths through ASTs~\cite{Alon2018,Alon2018a},
and embeddings for edits of code~\cite{Yin2018}.
Future work should study the impact of these representations on neural bug finding.

\subsection{Studies of Bug Finding Techniques}

% study of several techniques
A study related to ours applies different learning techniques to the bug detection problem~\cite{chappelly2017machine}.
Their data set includes seeded bugs, whereas we use real bugs.
Another difference is that most of their study uses manually extracted features of code, whereas we learn models fully automatically, without any feature engineering.
Their preliminary results with neural networks are based on a bit-wise representation of source code, which they find to be much less effective than we show token sequence-based models to be.

% studies that focus on recall
More traditional bug finding techniques have been subject to other studies, some of which focus on the recall of bug detectors~\cite{Thung2012,ase2018-study}, while others focus on their precision~\cite{Rutar2004,Wagner2005}.
The effectiveness of test generation techniques has been studied as well~\cite{Shamshiri2015,almasi2017industrial}.
Our work complements those studies by systematically studying neural bug finding.

\subsection{Defect Prediction and Unbalanced Data}

Machine learning models for software defect prediction~\cite{song2011general} suffer from data imbalance~\cite{wang2013using} (\ref{subsec:rq6}).
Skewed training data is usually tackled either by sampling techniques~\cite{kamei2007effects}, cost-sensitive learning~\cite{sun2007cost}, or ensemble learning~\cite{sun2012using}. Under-, over-, or synthetic-sampling techniques~\cite{kamei2007effects,bennin2018mahakil} have been applied to alleviate data imbalance in software defect prediction. 
Our approximated nearest neighbor (ANN) sampling of non-buggy examples (\ref{subsubsec:select_nonbuggy}) is a form of guided under-sampling.

\section{Conclusion}
\label{sec:conc}

This paper explores the opportunities and challenges of a novel way of creating bug detectors via deep learning.
We present neural bug finding and systematically study its effectiveness based on warnings obtained from a traditional static bug detection tool.
Studying neural bug detection models for 20 common kinds of programming errors shows that (i) neural bug finding can be highly effective for some bug patterns but fails to work well for other bug patterns, (ii) learned models pick up common code patterns associated with buggy code, as well as common instances of more general bug patterns, and (iii) surprisingly small data sets can yield effective models.
Our work also identifies some pitfalls associated with training and validating neural bug detectors and presents ways to avoid them.
We believe that this work is an important step into a promising new direction, motivating future work on more advanced neural bug finding tools and on improving the process of obtaining training data.

\bibliographystyle{IEEEtran}
\bibliography{references}

\end{document}